\documentclass[11pt]{article}
\usepackage{amssymb,times,graphicx}

\newcommand{\bE}{{\mathbb E}}

\newcommand{\bR}{{\mathbb R}}
\newcommand{\bZ}{{\mathbb Z}}

\newcommand{\cN}{{\cal N}}
\newcommand{\cO}{{\cal O}}

\newcommand{\cS}{{\cal S}}
\newcommand{\cT}{{\cal T}}
\newcommand{\cV}{{\cal V}}
\newcommand{\cU}{{\cal U}}
\newcommand{\cX}{{\cal X}}

\newcommand{\si}{\sigma}
\newcommand{\Ga}{\Gamma}
\newcommand{\veps}{\varepsilon}
\newcommand{\del}{\partial}
\newcommand{\pli}{\prod\limits}

\newcommand{\gtoas}[1]{{\quad\mathop{\longrightarrow}\limits_{#1}\quad}}
\newcommand{\abs}[1]{{\left\vert #1 \right\vert}}
\newcommand{\norm}[1]{{\left\Vert #1 \right\Vert}}
\newcommand{\sfrac}[2]{{\textstyle \frac{#1}{#2}}}

\newcommand{\I}{{\rm i}}
\newcommand{\E}{{\rm e}}
\newcommand{\dd}{{\rm d}}

\newcommand{\epkin}{\eta}
\newcommand{\eplo}{\varepsilon}
\newcommand{\ba}{{\bf a}}
\newcommand{\tnorm}[1]{|\!|\!| #1 |\!|\!|}

\newtheorem{theorem}{Theorem}[section]

\begin{document}

\title{Feynman graphs and renormalization \\ in quantum diffusion
\footnote{Talk given by M. Salmhofer at the conference in honour of 
Wolfhart Zimmermann's 80th birthday, Ringberg Castle, February 3--6, 2008}}

\author{L\'aszl\'o Erd\H os
\\[1ex]
\small
Mathematisches Institut, Universit\" at M\" unchen,
Theresienstr. 39, D-80333 M\" unchen\\
\small
\texttt{lerdos@mathematik.uni-muenchen.de}
\\[2ex]
Manfred Salmhofer
\\[1ex]
\small
Institut f\" ur Theoretische Physik, Universit\" at Leipzig,
Postfach 100920, 04009~Leipzig\\
\small
and Max--Planck--Institut f\" ur Mathematik, Inselstr.\  22,
D-04103 Leipzig \\ 
\small
\texttt{salmhofer@itp.uni-leipzig.de}
\\[2ex]
Horng--Tzer Yau
\\[1ex]
\small
Mathematics Department, Harvard University,
Cambridge, MA 02138, USA 
\\
\small
\texttt{htyau@math.harvard.edu}
}

\maketitle
\begin{abstract}
We review our proof that in a scaling limit, the time evolution 
of a quantum particle in a static random environment leads to
a diffusion equation.  In particular, we discuss the role of 
Feynman graph expansions and of renormalization.
\end{abstract}

\section{Introduction}
The emergence of irreversibility from reversible dynamics
in large systems has been one of the fundamental questions
in science since the days of Maxwell and Boltzmann. 
The famous debate about the statistical character
of the second law of thermodynamics and the related controversy
about Boltzmann's {\em Sto\ss zahlansatz} in the derivation of 
his transport equation has been very fruitful for physics and 
mathematics. After Lanford's rigorous justification of the Boltzmann 
equation for a classical many--particle system at short kinetic time 
scales\cite{Lanford}, the mathematical justification of the Boltzmann equation 
at longer timescales has remained a challenge up to the present time. 
The analogous statement for quantum systems remains open
even at the short kinetic timescale.

A related important question is to understand how Brownian 
motion emerges as an effective law from time-reversal-invariant 
microscopic physical laws, 
as given by a Hamiltonian system or the Schr\" odinger equation. 
Kesten-Papanicolaou\cite{KP} proved that the velocity distribution
of a classical particle moving in an environment consisting of
random scatterers (i.e., Lorenz gas with random scatterers)
converges to a Brownian motion in a weak coupling limit 
in dimensions $d\ge 3$.
In this model the bath of light particles whose fluctuations lead to 
the Brownian motion of the observed particle is
replaced with random static impurities.
A similar result was obtained in $d=2$ dimensions\cite{DGL2}.  
Recently\cite{KR}, the same evolution was controlled
on a longer time scale and the position process was proven 
to converge to Brownian motion as well.
Bunimovich and Sinai\cite{BS} proved the convergence of the
periodic Lorenz gas with a hard core interaction to a Brownian motion. 
In this model the only source of randomness is the distribution
of the initial condition.
Finally, D\"urr, Goldstein and Lebowitz\cite{DGL1} proved that
the velocity process of a
heavy particle in a light ideal gas, which is
a model with a dynamical environment, converges to the
Ornstein-Uhlenbeck process.

Although Brownian motion was discovered and first studied 
theoretically in the context of classical dynamics,  
it also describes the motion of
a quantum particle in a random environment, on a timescale that is long
compared to the standard kinetic timescale\cite{ESY1,ESY2,ESY3}. 
In the following we describe 
this result and the strategy of the proof in a bit more detail.  Besides the motivation
discussed above, the random Schr\" odinger operator that we study is also 
the standard model for transport of electrons in metals with impurities, 
which plays a central role in the theory of the metal--insulator transition\cite{Lee,VW}.
The outstanding open mathematical question in this area is the proof 
of the extended states conjecture, stating that in dimensions $ d \ge 3$,
at weak disorder, the spectrum of such Hamiltonians is absolutely continuous. 
Despite much effort, this conjecture has up to now only been proven\cite{Kl,ASW,FHS}
on the Bethe lattice, which can be interpreted as the case $d = \infty$.
In a system with a magnetic field, the existence of dynamical delocalization 
at certain energies near the Landau levels has been proven recently\cite{GKS}.

\section{The problem and the main result}
We consider random Schr\" odinger operators, 
both on a lattice and in the continuum, in $d \ge 3$
dimensions. In this presentation, we focus on the case $d=3$.
The time evolution of the {\em Anderson Model} (AM) is generated by 
\begin{equation}
\I \sfrac{\del}{\del t} \psi(t) = H \psi (t)
, \;
\psi (0) = \psi_0
\;
\mbox{ with }
\;
H =  -\frac12 \Delta + \lambda V_\omega
\;
\mbox{ on } 
\ell^2 (\bZ^d)
\end{equation}
where $-\Delta$ is the standard discrete Laplacian
and the potential is given by
$ V (x) = \sum_{a \in \bZ^d} V_{a} (x)$, with $ V_a(x) = v_a \delta_{x,a}$,
and $v_a$  independent identically distributed (i.i.d.) random variables.
We assume that
$m_k = \bE \left( v_a ^k \right)$ satisfies
\begin{equation}\label{momentcondition}
\forall i \le 2d: 
m_i < \infty , \qquad
m_1=m_3=m_5=0, \quad m_2 = 1.
\end{equation}
The continuum analogue of this model is the {\em Quantum Lorentz Model} (QLM), 
where $H =  -\frac12 \Delta + \lambda V_\omega$  on $L^2 (\bR^d)$, 
with $\Delta $ the standard Laplacian, $V_\omega (x) = \int_{\bR^d} B(x-y)\dd \mu_\omega(y)$,
where $B$ is a fixed spherically symmetric Schwarz function with $0 \in $ supp $\hat B$,
$\mu_\omega$ is a Poisson point process on $\bR^d$ with
homogeneous unit density and i.i.d.\ random masses:
\begin{equation}
\mu_\omega = \sum_{\gamma=1}^\infty v_\gamma(\omega) \delta_{y_\gamma(\omega)} .
\end{equation}
$\{ y_\gamma(\omega)\}$ is Poisson, 
independent of the weights $\{ v_\gamma(\omega)\}$.
Again, $m_k:=\bE_v \, v_\gamma^{k}$  is assumed to satisfy
(\ref{momentcondition}).

Suppose the initial state is localized, i.e.\ $\hat \psi_0$ is smooth.
How does the solution $\psi(t) = \E^{-\I t H} \psi_0$ behave
for large $t$ ?
If $\lambda =0$, the time evolution is easily calculated in 
Fourier space: $\hat\psi (t,k) = \E^{-\I t e(k)} \hat \psi_0 (k) $,
with $e(k) = k^2/2$ (QLM) or $e(k) = \sum_{i=1}^d (1- \cos k_i)$ (AM).
It is equally easy to see that the motion is ballistic, i.e.\ 
\begin{equation}
\langle X^2\rangle_{t} = \langle \psi(t) , X^2 \psi(t) \rangle 
\sim t^2 .
\end{equation} 
If $\lambda \ne 0$, one expects either localization, $ \langle X^2\rangle_{t}  = O(1)$ for all $t$, 
or diffusive behaviour (extended states), 
$\langle X^2\rangle_{t}  =  O(t)$, depending on $\lambda$ and $\hat \psi_0$. 
The localized behaviour corresponds to dense pure point spectrum at almost every energy;
this was proven for large disorder\cite{FS,AM} and away from the spectrum of the 
Laplacian. Extended states correspond to absolutely continuous spectrum. 
As mentioned, the latter has been proven\cite{Kl,ASW,FHS} on the Cayley tree
for small $\lambda > 0$.
At this time there is no proof of existence of extended states in $d=3$.
For a simpler case, namely that of randomness with a decaying envelopping function,
i.e.\  $V_\omega (x) = \omega_x h(x)$, $\omega_x$ i.i.d., $h$ fixed,  there is a proof\cite{RS,B}
that for $\eta > \frac12 $ and $h (x) \sim |x|^{-\eta}$ as $|x| \to \infty$, 
$H = - \Delta + V_\omega$ has absolutely continuous spectrum. 

Our result is formulated in terms of the {\em Wigner function}
\begin{equation}
W_\psi (x,v) 
=
\int \dd y \; 
\E^{\I vy} 
\overline{\psi (x+\frac{y}{2})} \, 
\psi (x-\frac{y}{2}) 
\end{equation}
which can be thought of as an analogue of a phase space density (but can become negative).
Its marginals are
$
\int W_\psi (x,v)  \dd x
=
|\hat \psi (v)|^2
$ and 
$
\int W_\psi (x,v)  \dd v
=
|\psi (x)|^2
$.
Moreover, 
\begin{equation}\label{hat-wig}
\hat W_\psi (\xi,v) = \int \dd x\; \E^{-\I x \xi} W_\psi (x,v)
=
\overline{\hat \psi (v - \xi/2)}\, \hat \psi (v+\xi/2) .
\end{equation}
On the lattice, one has to 
modify the definition of the Wigner transform slightly\cite{ESY3}.

The {\em kinetic scaling} is given by 
\begin{equation}
\epkin = \lambda^2, \qquad \cT = \epkin t, \qquad \cX= \epkin x,
\end{equation}
i.e.\ the microscopic time and space variables both become of order $\lambda^{-2}$,
so that velocities remain unscaled.

\begin{theorem}
\begin{equation}
\bE W^\epkin_{\psi(\cT \epkin^{-1})}  (\cX,\cV)
\gtoas{\epkin \to 0} F(\cX,\cV,\cT),
\end{equation} 
$F$ the solution of the {\em linear Boltzmann equation}
\begin{eqnarray}
& &\frac{\del}{\del \cT} F(\cX,\cV,\cT) 
+ 
(\nabla e) (\cV) \cdot \nabla_{\cX} F(\cX,\cV,\cT)
\nonumber\\
&&=
2\pi \int \dd \cU \;
\delta(e(\cU)-e(\cV)) \;
\abs{\hat B(\cU - \cV)}^2 \;
\left[
F(\cX,\cU,\cT) - F(\cX,\cV,\cT)
\right] .
\end{eqnarray}
\end{theorem}

\noindent
This theorem was first proven for the continuum for small
time $\cT$ \cite{Sp1}, then for arbitrary time\cite{EY},
and later extended to the lattice case\cite{Ch}.

The {\em diffusive scaling} is defined by 
\begin{equation}
\eplo = \lambda^{2+\kappa/2},
\qquad
X = \eplo x,
\qquad
T= \eplo \lambda^{\kappa/2} t = \lambda^{\kappa+2} t .
\end{equation}
This is long compared to the kinetic timescale: 
the kinetic variables $\cX $ and $\cT$ diverge as $\lambda \to 0$
when $X$ and $T$ are kept fixed,
\begin{equation}
\cX = \lambda^{-\kappa/2} X , 
\qquad
\cT = \lambda^{-\kappa} T.
\end{equation}
A first hint at diffusion is that under this scaling $\cX^2 / \cT =X^2/T$ 
is independent of $\lambda$. The result for the Anderson model is

\begin{theorem} 
Let $d=3$, $\psi_0 \in \ell^2(\bZ^3)$ and $\psi(t)$ be the solution to the 
random Schr\" odinger equation with initial condition $\psi_0$. 
If  $\kappa > 0$ is small enough
and $\eplo = \lambda^{2+\kappa/2}$, 
then in the limit $\lambda  \to 0$,
$\bE W^\eplo_{\psi(\lambda^{-2-\kappa }T)} $ converges
weakly to the solution $f$ of a heat equation.

More precisely: denote 
$
\langle F\rangle_E 
=
\Phi(E)^{-1}
\int \dd v\; F(v) \delta(E-e(v)) ,
$
where $ \Phi(E) = \int \dd v\; \delta(E-e(v))$.
Let $E \in (0,3)$  and 
$
D_{ij} (E) 
=
\frac{1}{2\pi \Phi(E)}
\langle
\nabla_i e\; \nabla_j e
\rangle_E,
$
and let $f$ be the solution of the heat equation
\begin{eqnarray}
\frac{\del}{\del T} f(T,X,E) 
&=& 
\nabla_X \cdot D(E) \nabla_X\;  f(T,X,E) 
\\
f(0,X,E)
&=&
\delta(X) \; 
\langle
|\hat\psi_0|^2 \rangle_E \; .
\end{eqnarray}
Let $\cO(x, v)$ be a Schwartz function on $\bR^d\times \bR^d/2\pi \bZ^d$. Then
\begin{eqnarray}
&&
\lim_{\eplo \to 0} 
\sum\nolimits_{X \in (\eplo\bZ/2)^d} 
\int \! \dd v \; \cO(X, v)  
\; \bE
 W^\eplo_{\psi(\lambda^{-\kappa-2} T)} (X, v)
\nonumber\\
&&
\qquad = \int_{\bR^d} \dd X \int \!\! \dd v \; \cO(X, v) \; f(T, X, e(v))  .
\end{eqnarray}
The limit is uniform on $[0,T_0]$ for any $T_0 > 0$.
\end{theorem}

\noindent
We discuss some of the ideas in the proof of this theorem in Section \ref{proofsec}.

\medskip\noindent
If $\hat \psi_0 \in C^1$ 
and $\lambda $ is small enough, we have the more detailed error estimate
\begin{eqnarray}
&&
\int \dd v \int \dd \xi \; \hat \cO (\xi,v) \; 
\bE \hat W^\eplo_{\psi (\lambda^{-2-\kappa} T)} (\xi,v)
\\
&=&
\int \dd \xi \int \Phi (E) \dd E \;
\E^{- \frac{T}{2} \langle \xi, \; D(E) \xi \rangle_E}
\langle \hat \cO (\xi, \cdot ) \rangle_E\;
\langle \hat W_{\psi_0} (\eplo \xi, \cdot )\rangle _E
+
o(1) .
\nonumber
\end{eqnarray}
The Boltzmann equation also gives the same diffusion equation 
in the long time limit, but it was itself derived from the
quantum mechanical time evolution only for shorter timescales.
The main difficulty in the proof is to deal with contributions that
vanish for $\lambda \to 0$ under kinetic scaling, but that become
important under the above--defined diffusive scaling. 
More technically speaking, in the Feynman expansion done 
to analyze the time evolution, most of these terms would even diverge
under diffusive scaling if we did not renormalize the propagator. 

The allowed values of $\kappa $ are in an interval  $[0,\kappa_0)$,
where $\kappa_0$ is a universal constant. 
For technical  reasons, $\kappa_0 $ has to be chosen very small 
in the proof. Heuristically, i.e.\ ignoring many of the technical 
complications and assuming optimal bounds, one would expect
the remainder of the renormalized Feynman graph expansion to vanish up to
$\kappa_0=2$, and to diverge for $\kappa_0 > 2$. 

The diffusive scaling leads to a diffusion on the energy shells. 
A diffusion mixing energy shells is expected to start at $t = \lambda^{-4}$.

An intuitive way of interpreting the expansion described below
is as an expansion in the number $N$ of collisions of the particle 
with the randomly (but statically) arranged obstacles represented 
by the potential. As compared to the previous results\cite{EY,Ch},
the main new feature here is that under diffusive scaling, 
the effective number of collisions of the  particle diverges. 
That is, not only is it necessary to expand to an order $N$ that
diverges as $\lambda \to 0$, but also the main contribution 
does not come from terms with a finite number of collisions. 

\section{Collision histories, Feynman graphs, and ladders}\label{proofsec}
We discuss some of the ideas of the proof for the example 
of the Anderson model,  i.e.\ the lattice situation. 
For the detailed bounds of Feynman graphs, the lattice
leads to a number of complications\cite{ESY3}, but for the presentation
it is easier. 

\subsection{Collision histories}
Let us start with a formal time--ordered expansion, setting 
$H_0 = - \frac12 \Delta$ and expanding in $\lambda V$. 
Then 
$\psi(t) = \E^{-\I t H} \psi_0 = \sum_{n \ge 0} \psi^{(n)} (t)$
with
\begin{eqnarray}
 \psi^{(n)} (t) 
&=& 
(-\I \lambda)^n
\int \dd \mu_{n+1} (s)
\E^{-\I s_{n} H_0} V \E^{-\I s_{n} H_0} \ldots 
V \E^{-\I s_{0} H_0} \psi_0 
\end{eqnarray}
where $s= (s_0, \ldots, s_n)$ and
\begin{equation}\label{mudef}
\dd \mu_{n+1} (s)
=
\int\limits_{[0,\infty)^{n+1}}
\dd s_0 \ldots \dd s_n \;
\delta \left( t - \sum_{j=0}^{n} s_j\right) .
\end{equation}
Because $V=\sum_{a\in \bZ^d} V_a$,  it is natural to split
each $\psi^{(n)}$ further,  
$\psi^{(n)} (t) = \sum_{\ba_n} \psi^{(n)}_{\ba_n} (t)$.
Every sequence  of obstacle labels 
$\ba_n = (a_1,\ldots,a_n) \in (\bZ^d)^n$
represents a {\em collision history}, 
and for $k \in \{1, \ldots, n-1\}$, the time variables $s_k$ in (\ref{mudef}) 
are the time differences between two subsequent collisions.  
The delta function in (\ref{mudef})
enforces the constraint that these time differences, together with the 
propagation times $s_0$ before the first collision and $s_{n}$ after the 
last one, add up to the total time $t$. We shall discuss convergence 
questions about this expansion later.

Our detailed analysis takes place in momentum space, where each
$V$ acts as a convolution operator, so that 
\begin{eqnarray}\label{psimom}
\hat \psi_n (t,p_n)
\!=\!
(\mbox{-}\I)^n\!
\int \dd \mu_{_{n+1}} \!(s)\!
\int \!
\pli_{j=0}^{n-1} \sfrac{\dd^d p_j }{(2\pi)^d} \, \E^{-\I s_j e(p_j)} \!
\pli_{j=1}^n \hat V (p_j \mbox{-} p_{j-1})
\hat \psi_0 (p_0) .
\end{eqnarray}
Very schematically, one can represent this 
as follows, where each of the dashed lines represents
a factor $\lambda V$ and each of the full lines 
gets a phase factor $ \E^{-\I s_j e(p_j)}$.
\begin{center}
    \includegraphics[height=10mm]{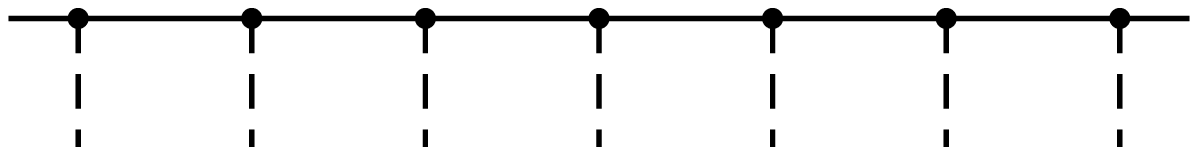}
\end{center}
As the convolution formula shows, one can define a momentum flow
in this graph, where the momentum change $p_j - p_{j+1}$ 
flows away through the dashed line. Before the disorder average, 
there is no translation invariance in the system, so every scattering
at an obstacle changes the momentum of the particle. 

\subsection{Disorder average and graphs}

Recalling (\ref{hat-wig}), we have 
\begin{equation}
\bE \left[\hat W_{\psi(t)} (\xi,v)\right]
=
\sum_{n,n'} \sum_{\ba_n,\ba'_{n'}} 
\bE \left[
\overline{\hat \psi^{(n)}_{\ba_n} (t,v - \xi/2)}
\hat \psi^{(n')}_{\ba'_{n'}} (t,v+\xi/2) 
\right] .
\end{equation}
Note that there are now two, a priori independent, collision histories, 
one for $\psi$ and one for $\bar \psi$. It will be part of the proof 
to show that, in the scaling limit we consider, the only contributions
after self--energy renormalization come from the so-called ladder graphs, 
where the two collision histories are identical: $n=n'$  and ${\bf a}_n = {\bf a}'_n$.

Because the disorder is i.i.d., translation invariance holds for the 
average, which means that momentum conservation also holds
for the dashed lines, which for the Anderson model simply 
correspond to a factor $\lambda^2$, since the second moment 
of the disorder was normalized to $1$ in (\ref{momentcondition}).

The result can be represented as a graph built of two particle
lines, particle--disorder vertices, which are joined by disorder lines, 
and, if the randomness is non-Gaussian, 
disorder-disorder vertices, which correspond to the higher 
moments of the disorder distribution. An example is 
\begin{center}
    \includegraphics[height=20mm]{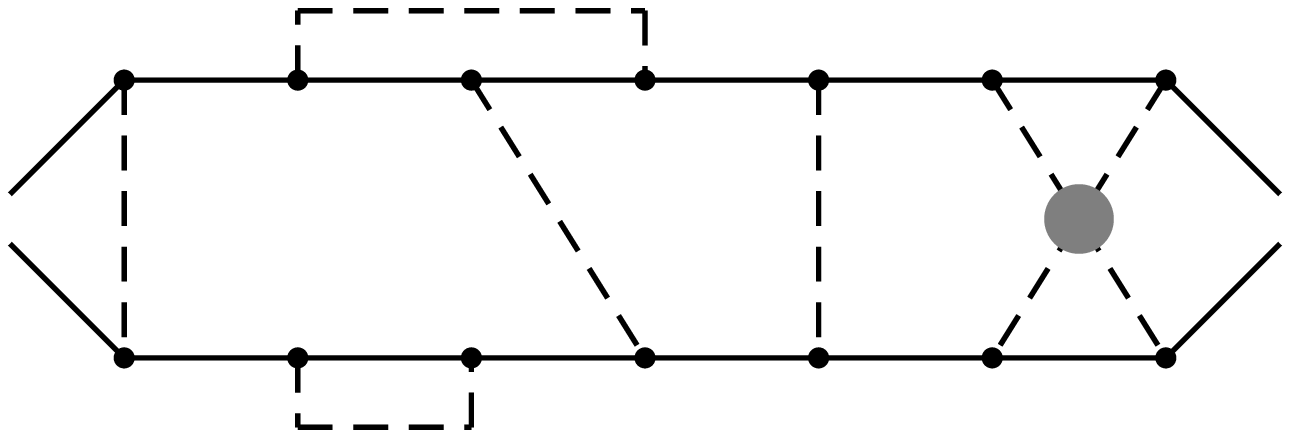}
\end{center}
Particle lines get propagators $\E^{-\I s_j e(p_j)}$, interaction lines
give factors $\lambda^2$, and the disorder-disorder vertex
of degree four corresponds to a factor $m_4 \lambda^4$.

It is clear that in the way the expansion was introduced above, 
one really needs the assumption that arbitrary moments, not just the first 
$2d$ ones, exist.  The expansion employed in the true proof contains a 
stopping rule which avoids high moments, but we shall not discuss this
here in more detail. In fact, we shall in the following assume 
for simplicity that the disorder is  Gaussian, so that there are no 
vertices of higher degree for the dashed lines, and the average just
corresponds to a  pairing of interaction lines.

An example of a pairing is as follows
\begin{center}
    \includegraphics[height=20mm]{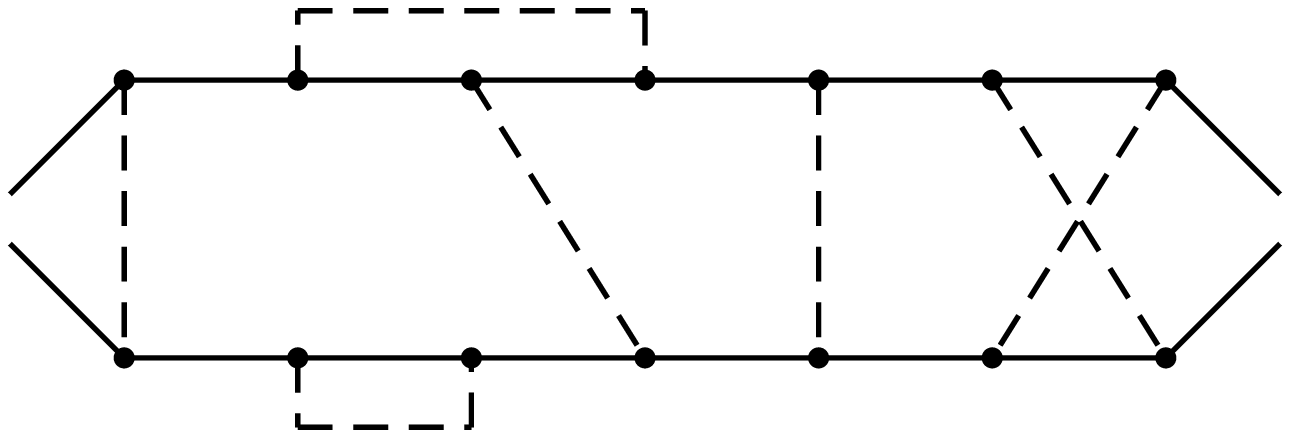}
\end{center}
Note that here, there is  a crossing of the two pairing lines 
in the graphical representation, but there are no vertices in which more than
one interaction line enters.

A special class of pairings are the {\em up--down pairings}, 
where $n=n'$ and the pairing corresponds to a permutation $\si \in \cS_n$:
\begin{center}
    \includegraphics[height=20mm]{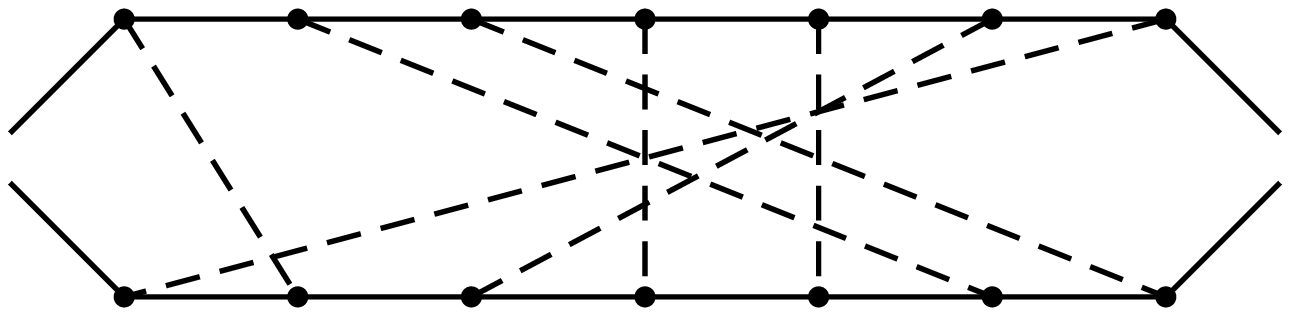}
\end{center}

The most important term turns out to be the {\em ladder graph}, 
corresponding to $\si = \mbox{ id:} $ 

\begin{center}
    \includegraphics[height=20mm]{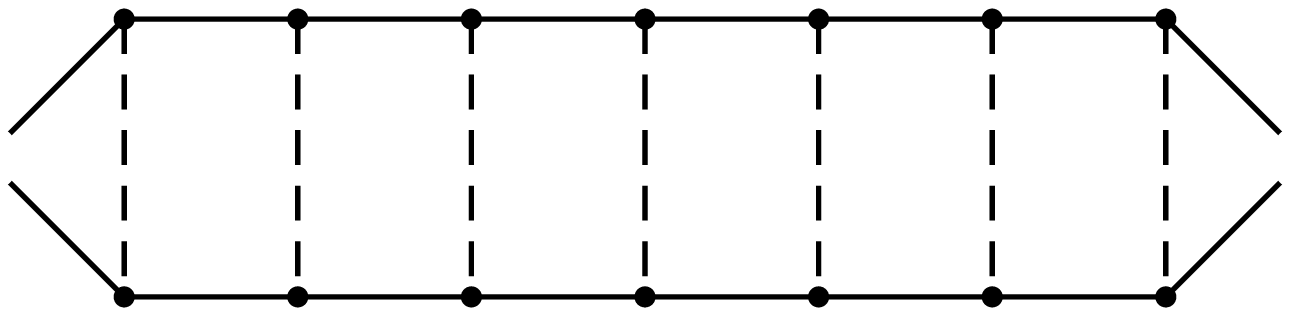}
\end{center}

\subsection{Graph bounds}
In the following, we give a brief discussion of bounds of the 
contributions of individual graphs,  restricting to up--down pairings. 
If one takes a bound in the representation (\ref{psimom}), 
each phase factor is replaced by $1$.
This leads to a bound of order $(\lambda t)^n/n!$ (where the $n!$ comes from 
the time ordering implied by the delta function in (\ref{mudef})), 
which does not even allow to consider the kinetic scaling where
$\lambda^2 t$ is fixed.  For this reason, the following 
{\em propagator representation} is useful. Let $\eta > 0$. 
Then, inserting the Fourier representation of the delta function,
\begin{eqnarray}
&&
\int_{[0,\infty)^{n+1}} \dd^{n+1} s\; 
\delta \left( t - \sum_{j=1}^{n+1} s_j \right) 
\prod\limits_{j=1}^{n+1}
\E^{-\I s_j e(p_j)}
\nonumber
\\
&=&
\E^{t\eta}
\int_{[0,\infty)^{n+1}} \dd^{n+1} s\; 
\delta \left( t - \sum_{j=1}^{n+1} s_j \right) 
\prod\limits_{j=1}^{n+1}
\E^{-\I s_j (e(p_j) - \I \eta)}
\nonumber
\\
&=&
\E^{t\eta}
\int \sfrac{\dd\alpha}{2\pi} \;
\E^{-\I t \alpha}
\int_{[0,\infty)^{n+1}} \dd^{n+1} s\; 
\prod\limits_{j=1}^{n+1}
\E^{-\I s_j (\alpha - e(p_j) + \I \eta)}
\nonumber
\\
&=&
\I^{-n}
\E^{t\eta}
\int \sfrac{\dd\alpha}{2\pi} \; 
\E^{-\I \alpha t}
\prod\limits_{j=1}^{n+1}
\frac{1}{\alpha - e(p_j) + \I \eta} \; .
\end{eqnarray}
It is convenient to choose $\eta = t^{-1}$. 

The contribution of a permutation $\sigma \in \cS_n$,
corresponding to an up--down pairing graph $\Gamma_\sigma$,
to $\langle \hat \cO , \hat W^{\veps}_{\psi} \rangle $ is
\begin{eqnarray}
Val(\Ga_\si)
&=&
\lambda^{2n} \; \E^{2t\eta}
\int \sfrac{\dd \alpha \, \dd \beta}{(2\pi)^2}\;
\E^{\I (\beta - \alpha) t}
\nonumber
\\
&&
\int \dd \xi \int \pli_{j=0}^n \sfrac{\dd^d p_j }{(2\pi)^d}\int \pli_{k=0}^n \sfrac{\dd^d q_k }{(2\pi)^d} \;\;\;
\hat\cO (\xi,p_n) \hat W^{\veps}_{\psi_0} (\xi,p_0)
\nonumber
\\
&&
\pli_{j=0}^n
\frac{1}{\beta - \overline{\omega(q_j - \frac{\veps \xi}{2})} - \I \eta}\;
\frac{1}{\alpha - \omega(p_j - \frac{\veps \xi}{2}) - \I \eta}
\nonumber
\\
&&
\pli_{j=1}^n \delta \left( p_j - p_{j-1} - (q_{\si(j)} - q_{\si(j) -1}) \right) .
\end{eqnarray}
At the moment, $\omega (p) = e(p) \in \bR$; later,
$\omega$ will change under renormalization
and become complex.

A simple Schwarz inequality separating the dependence on the $p_i$ and 
that on the $q_i$ implies that for all $\si$
\begin{equation}
\abs{Val(\Ga_\si)} \le Val(\Ga_{id}) .
\end{equation}
The ladder is easy to calculate at $\xi = 0$, and 
a ladder of length $n$
is of order $\frac{1}{n!}(\lambda^2 t)^{n} = \frac{1}{n!} \cT^n$. 

A crucial observation is that the values of graphs with {\em crossings} 
get inverse powers of $t$, as compared to the ladder. 
This follows from the bound
\begin{equation}
\int \dd p\;
\frac{1}{|\alpha - \omega(p) + \I \eta|}\;
\frac{1}{|\beta - \overline{\omega(\pm p + q)} - \I \eta|}
\le
C |\log \eta|^3
\frac{\eta^{-b}}{\tnorm{q} + \eta}
\end{equation}
($b=0$ for the continuum; $1/2 \le b \le 3/4$ on the lattice).
$\tnorm{p} = |p|$ {in the continuum}, 
$\tnorm{p} = \min\{ |p-v|: v_i \in \{0, \pm \pi\}\}$ on the lattice.
Again, here $\omega(p) = e(p)$.
This motivates why the ladder graph gives the dominant contribution 
under kinetic scaling. 
However, the number of graphs goes like $n!$, which cancels the $1/n!$, 
hence expanding to infinite order one gets majorants by geometric series, 
which converge  only on very short kinetic timescales $\cT$.
This is the reason for the restriction to small kinetic timescales 
in the first proof\cite{Sp1} of the Boltzmann equation for the QLM.

\subsection{Expansions to finite order and remainder terms}

Major progress\cite{EY} came from the realization that 
one can do an expansion to finite order with an efficient 
remainder estimate. 
A natural way to generate a finite--order expansion is the {\em Duhamel formula}
\begin{equation}
\psi(t) = \E^{-\I t H} \psi_0 
=
\E^{-\I t H_0} \psi_0 + \int_0^t \dd s\; 
\E^{-\I (t-s) H} \lambda V \E^{-\I s H_0} \psi_0 .
\end{equation}
Iteration gives
\begin{equation}
\psi(t) = \sum_{n=0}^{N-1} \psi^{(n)} (t) + \Psi_N (t),
\end{equation}
where
\begin{equation}
\Psi_N (t) = (-\I) \int_0^t \dd s\; \E^{-\I (t-s) H} \lambda V \psi^{(N-1)} (s)
\end{equation}
and
\begin{equation}
\psi^{(n)} (t) 
=
(-\I \lambda)^n
\int \dd \mu_{n+1} (s)
\E^{-\I s_{n} H_0} V \ldots V\; \E^{-\I s_0 H_0} \psi_0 .
\end{equation}
An alternative way of looking at this is via 
its relation to the resolvent formula
\begin{equation}
R_z 
=
R_z^{(0)}
+ R_z \lambda V R_z^{(0)}
\end{equation}
where $R_z = (z-H)^{-1}$ and $R_z^{(0)} = (z-H_0)^{-1}$.
Iteration of the resolvent equation and using the Fourier 
transform gives the above propagator representation directly. 
The Duhamel formula is obtained via
\begin{equation}
\E^{-\I t H } 
=
- \E^{t \eta} 
\int
\sfrac{\dd \alpha}{2\pi \I} \; 
\E^{- \I \alpha t} \;
R_{\alpha + \I \eta} .
\end{equation}

\noindent
The second crucial ingredient is that one can use
the unitarity of the full time evolution to reduce all terms 
to ones where no $H$ appears in the 
time evolution any more:
\begin{eqnarray}
\norm{\Psi_N (t)} 
&\le &
\int_0^t \dd s\; \norm{\E^{-\I (t-s) H} \lambda V \psi^{(N-1)} (s)}
\le
\int_0^t \dd s\; \norm{ \lambda V \psi^{(N-1)} (s)} .
\end{eqnarray}
Thus 
\begin{equation}
\norm{\Psi_N (t)} ^2
\le t\;  |\lambda|^2
\int_0^t \dd s\; \norm{ V \psi^{(N-1)} (s)}^2.
\end{equation}
The remaining integral over $s$ effectively gives a factor $t$, 
which is the price to pay for this unitarity bound. 
To control this factor, one needs exhibit more factors $t^{-1}$ in graphs 
with several independent crossings, and treat graphs with 
only one crossing explicitly (in the resolvent iteration, the 
unitarity bound would be replaced by $\norm{R_{\alpha + \I \eta} } \le \eta^{-1}$).

By a Schwarz inequality, one can see that the Wigner transform is 
continuous in $L^2$ norm: 
\begin{equation}
\abs{\bE 
\left(
\langle \hat \cO , \hat W^\veps_{\psi_1} \rangle
-
\langle \hat \cO , \hat W^\veps_{\psi_2} \rangle
\right) }
\le
C
\int\dd \xi \; \sup\limits_{v} \abs{\hat \cO (\xi,v)} \;
\sqrt{
\bE \norm{\psi_1}^2 \; 
\bE \norm{\psi_1-\psi_2}^2
} .
\end{equation}
Thus the unitarity bound can also be used for the Wigner transform. 
The proof  of the Boltzmann equation\cite{EY} on an arbitrarily large
kinetic timescale $\cT$  uses an expansion up to order
$N \sim \log t$. 
The ladder terms give the gain  term in the Boltzmann equation. 
The lowest order self--energy  correction gives the loss term
in the Boltzmann equation.  It corresponds to the ``gate'' graph
\begin{center}
    \includegraphics[height=15mm]{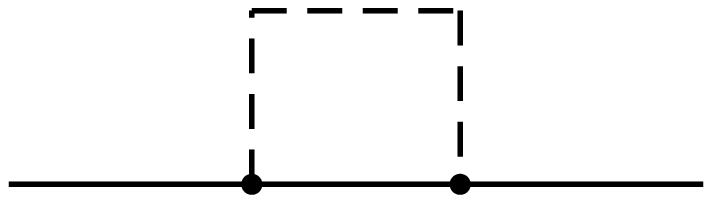}
\end{center}

\subsection{Long time scale: renormalization}
Because the ladder with $n$ rungs is of order $ (\lambda^2 t)^n / n! $,
it diverges under diffusive scaling, and so do other graphs. 
To increase the time beyond $\lambda^{-2}$, we need to do 
a renormalization. Formally, one can think of this as a resummation 
of the gate diagrams, which are of self--energy type, but this geometric 
series converges only for small $\lambda^2 t$. A way to avoid such 
formal resummations is to change the way $H$ is split into a ``free''
and an interaction part, i.e., expand around a different $H_0$.
For $\veps > 0 $ set
\begin{equation}
\Theta_\veps (\alpha) = \int \dd q \frac{1}{\alpha - e(q) + \I \veps} \; .
\end{equation}
This is the value of the gate diagram at energy $\alpha$ in the Anderson model 
(in the QLM, the integrand contains an additional factor from the interaction function).
The limit $\Theta (\alpha) = \lim_{\veps \to 0+} \Theta_\veps (\alpha)$ exists
and is H\" older continuous\cite{ESY1} in $\alpha$ of order $1/2$.
Let
\begin{equation}
\theta(p) = \Theta (e(p)) .
\end{equation}
The idea is now to put $\lambda^2 \theta (p)$ as a counterterm, 
which by construction subtracts every insertion of a gate diagram at the point 
$\alpha = e(p)$ where the particle propagator is singular. Because
$\alpha$ and $\eta$ appear only as auxiliary quantities in the expansion, 
it was necessary to take $\veps \to 0$ above and define $\theta$ in an 
$\alpha$--independent way. 

The counterterm is added and subtracted so that the Hamiltonian does not change:
let $\omega (p) = e(p) + \lambda^2 \theta(p) $ and decompose
\begin{equation}
H= \omega (P) + U, \qquad U = \lambda V - \lambda^2 \theta(P)
\end{equation}
(where $P$ denotes the momentum operator). 
The function $\omega$ can be thought of as a new dispersion relation of
energy as a function of momentum. However, $\omega$ also 
has a negative imaginary part, roughly of order $\lambda^2$. 
More precisely, for $d \ge 3$ there is $c >0$ such that
\begin{equation}
\mbox{ Im } \omega (p) \le - c \lambda^2 \tnorm{p}^{d-2} .
\end{equation}
Thus $H_0$ is no longer selfadjoint. However, the negative sign 
of Im $\omega$ implies that the resolvent $R_{\alpha + \I \eta}$
is still well-defined, since the imaginary parts add up with the same 
sign. Correspondingly, the time evolution operator $\E^{-\I s H_0}$ is 
no longer unitary but it remains bounded for $s \ge 0$.
Both the Duhamel and the resolvent iteration are thus well-defined.
Besides the new propagator $(\alpha + \I \eta - \omega (p))^{-1}$, 
the important change is that every factor $U$ now also contains 
a counterterm insertion $- \lambda^2 \theta (p)$. 
The point about these iterations is that they can be stopped (or even
modified) after every expansion step. It is thus clear that one can 
group the counterterms that appear in the expansion together 
with the gates that get created when taking the average over the 
disorder.  The cancellation among these two terms provides a small 
factor that makes such terms vanish in the diffusive scaling limit. 
Moreover, it is clear that one can implement rules for 
stopping the expansion independently of the subsequent disorder average. 
In particular, because the randomness is i.i.d., one can avoid moments
beyond the power $2d$ by stopping the expansion when a given site
has appeared in the collision history $d$ times. 
The terms to which no such repetition or renormalization cancellation
applies are expanded up to order 
$n \sim \lambda^2 t  \lambda^{-\delta} \sim \lambda^{-\kappa-\delta} $, 
where $\delta > 0$ depends on $\kappa$. The intuition behind this is that
certain graphs with $n \sim \lambda^2 t \sim \lambda^{-\kappa}$ give the 
main contribution, and expanding up to an order that is $\lambda^{-\delta}$ 
higher leads again to small factors. 

The imaginary part of $\omega$ gives effectively a regularization 
$O(\lambda^2)$ instead of $O(\eta)$ for the denominators, 
which changes the values of all diagrams significantly. 
In particular, the integral for one rung of the ladder becomes
\begin{equation}
 \int \sfrac {\lambda^2\; \dd p}{ (\alpha - \overline{\omega(p+r)}
      -i\eta)
     \; (\beta - \omega(p-r)  +i\eta)} \;
=     
 1+ C_0\lambda^{1- O(\kappa)}\; .
\end{equation}
where $C_0$ is a constant. 
Thus with this renormalization, the ladders become of order 1,
so that one can go beyond kinetic scaling. 
Indeed, in the language of Feynman graphs, the main result 
can be stated informally as

\begin{quote}
After renormalization, the sum of the ladder graphs for the Wigner
transform converges to a solution of the heat equation in the 
diffusive scaling limit. 
\end{quote}

The precise statements are Theorems 5.1, 5.2, and 5.3 in Ref.\cite{ESY1}.
They involve in particular proving that the terms which do not 
correspond to pure up--down pairings vanish in the limit, 
and dealing with a number of technical complications which 
arise from the fact that one has to do an expansion to a finite order. 

\subsection{The key estimate for controlling combinatorics}
We have had to leave out almost all technical details
to avoid overloading the presentation, 
but we should like to at least mention 
the heart of the proof  here at the end, 
to clarify the main ideas about the Feynman graph expansion. 

Focusing on up-down pairings, 
we have to deal with a combinatorial problem of 
bounding the sum over the $n!$ permutations $\sigma \in \cS_n$. 
As mentioned, with an expansion to infinite order, 
one cannot get beyond the kinetic scaling because of this
factor $n!$. The control of the remainders is done here
by choosing  an appropriate stopping $n$ for the expansion
and by ``beating down the combinatorics by power counting''. 
That is, we prove exponential
suppression of the values of Feynman graphs in the 
number of crossings they have, that is, loosely speaking, 
in their complexity. 

The precise notion capturing the complexity of a permutation
$\sigma \in \cS_n$ is its degree $d(\sigma)$, defined as  
the number of non--ladder and non--antiladder indices. 
Essentially, the ladder indices are those for which $\sigma (i+1) = \sigma (i) + 1$, 
and the antiladder indices are those for which $\sigma(i+1) = \sigma (i) -1$. 

\begin{theorem}
Let $\Ga_\si$ be the Feynman graph corresponding to
$\sigma$. There is $\gamma > 0$ such that for all $\sigma$
\begin{equation}\label{combest}
\abs{Val(\Ga_\si)} \le C \lambda^{\gamma d(\sigma)} .
\end{equation}
\end{theorem}

\noindent
This theorem is proven using a special integration algorithm for 
bounding the values of large Feynman graphs\cite{ESY1}. 

The number of permutations with degree $D$ is 
\begin{equation}
\cN_{n,D} 
=
|\{ \sigma \in \cS_n : d(\sigma) =D \}|
\le 2 (2n)^D .
\end{equation}
Expanding up to $n = O(\lambda^{-\kappa - \delta})$, $\delta >0$, we have
by (\ref{combest}), 
{if $\gamma - \kappa - \delta > 0$},
\begin{equation}
\sum_{{\sigma \in \cS_n \atop d(\sigma) \ge D}}
\lambda^{\gamma  d(\sigma)}
=
\sum_{d=D}^k \lambda^{\gamma d} \cN_{n,d}
\le
2
\sum_{d=D}^k (2\lambda)^{d(\gamma - \kappa - \delta)}
\le 
O(\lambda^{D (\gamma - \kappa - \delta)}).
\end{equation}
Thus the contribution from the sum of all terms with degree $D \ge 2$ 
is small if $\gamma - \kappa - \delta > 0$, hence the essential restriction for 
the value of $\kappa $ is that of $\gamma$. As mentioned, one would
hope to get close to $\gamma=2$ in (\ref{combest}), 
but $\gamma $ has to be chosen smaller for technical reasons.

\section{Conclusion}
We have shown that, for random Schr\" odinger operators
with a weak static disorder
the quantum mechanical time evolution can be controlled
on large space and time scales where a diffusion equation
governs the behavior. The Schr\"odinger evolution is time--reversible --
yet irreversibility on large scales emerges. This
apparent controversy is resolved by noting that along the scaling limit
microscopic degrees of freedom have been effectively integrated out.

Although the expansion methods we use bear some resemblance to those
of constructive quantum field theory, there are also a few
noteworthy differences.
First, because we analyze the time evolution at real time,
the (near--)singularities of the propagators
are located on hypersurfaces, and not at points, as would be the case
in Euclidean field theories.
The singularity structure is to some extent similar to that in
real time Fermi surface problems, although  there is no
fixed Fermi surface here -- the integrals
over $\alpha$ and $\beta$ ``test'' all possible level
sets of the function $e(p)$, and this leads to a number
of serious complications.
Second, we are able to control the combinatorics of a
straightforward Feynman graph expansion in momentum space,
while the analysis in constructive field theory (to our knowledge,
always) needs to be done by cluster expansions in position space
to avoid the divergence of an infinite series of Feynman graphs.
The reason for this is twofold:
the unitarity bound allows us to do an expansion to a finite order,
and our strong improvement (\ref{combest}) over standard power counting bounds allows us to push this order so high that we can reach the scale
where diffusion sets in, while still retaining control of the remainders.

The genuine challenge is to show diffusion without taking scaling limits,
i.e. for a fixed (small) disorder $\lambda$ and for any time independent
of $\lambda$. With expansion techniques, this would require
to renormalize not only the self--energy to arbitrary order but
also the four--point functions. Refining the self--energy renormalization
poses no fundamental difficulty. The correct renormalization of all
four--point functions in this problem, however, remains a widely open problem. 

\bibliographystyle{ws-procs9x6}

\begin{thebibliography}{99}

\bibitem{Lanford}
O.E.\ Lanford III,{\sl On the derivation of the Boltzmann equation.}
Ast\'erisque {\bf 40}, 117-137 (1976)

\bibitem{KP}
H.  Kesten, G. Papanicolaou: {\sl
A limit theorem for stochastic acceleration.}
Comm. Math. Phys. {\bf 78}  19-63. (1980/81)

\bibitem{KR} T. Komorowski, L. Ryzhik: {\sl Diffusion in a weakly random
Hamiltonian flow.} 
Commun. Math. Phys. {\bf 263} no.2. 277-323 (2006)

\bibitem{DGL2} D. D\"urr, S. Goldstein, J. Lebowitz:
{\sl Asymptotic motion of a classical particle in random potential
in two dimensions:  Landau model}, Commun. Math. Phys. {\bf 113} (1987) no 2.
209-230.

\bibitem{BS} L. Bunimovich, Y. Sinai:
{\sl Statistical properties of Lorentz
gas with periodic configuration of scatterers. }
Commun. Math. Phys.  {\bf 78} no. 4, 479--497 (1980/81),

\bibitem{DGL1} D. D\"urr, S. Goldstein, J. Lebowitz:
{\sl A mechanical model of Brownian motion.}
Commun. Math. Phys. {\bf 78} (1980/81) no. 4, 507-530.

\bibitem{ESY1}  L. Erd\H os, M. Salmhofer and  H.-T. Yau,
{\sl  Quantum diffusion of the random Schr\"odinger evolution in the
scaling limit.} 
Advances in Mathematics  (2008) 
DOI  10.1007/s11511-008-0027-2

\bibitem{ESY2}  L. Erd\H os, M. Salmhofer and  H.-T. Yau,
{\sl  Quantum diffusion of the random Schr\"odinger evolution in the
scaling limit II.  The recollision diagrams.} 
Commun. Math. Phys. {\bf 271}, 1-53 (2007)

\bibitem{ESY3}  L. Erd\H os, M. Salmhofer and  H.-T. Yau,
{\sl Quantum diffusion for the Anderson model in
scaling limit.} Ann. Inst. H. Poincare {\bf 8}, 621-685 (2007)

\bibitem{Lee} P. A. Lee, T. V. Ramakrishnan, {\sl Disordered
electronic systems. \/} Rev. Mod. Phys. {\bf 57}, 287--337 (1985)

\bibitem{VW} D. Vollhardt, P. W\"olfle, {\sl Diagrammatic,
self-consistent treatment of the Anderson localization problem
in $d\leq 2$ dimensions. \/} Phys. Rev. B {\bf 22}, 4666-4679
(1980)

\bibitem{Kl}
A. Klein, {\sl
Absolutely continuous spectrum in the Anderson model on the Bethe
lattice}, Math. Res. Lett. {\bf 1}, 399--407 (1994)

\bibitem{ASW} M. Aizenman, R Sims, S. Warzel, {\sl Absolutely continuous
spectra of quantum tree graphs with weak disorder.}
Commun. Math. Phys. {\bf 264} no. 2, 371-389 (2006)

\bibitem{FHS} R. Froese, D. Hasler, W. Spitzer,
{\sl Transfer matrices, hyperbolic geometry and absolutely 
continuous spectrum for some discrete Schr\" odinger operators on graphs.}
J. Funct. Anal. {\bf 230} no 1, 184-221 (2006)

\bibitem{GKS}
F.\ Germinet, A.\ Klein, J.\ Schenker, 
{\sl Dynamical delocalization in random Landau Hamiltonians}.
Ann. Math. {\bf 166} (2007) 215 -- 344

\bibitem{FS}
J. Fr\"ohlich and T. Spencer,
{\sl Absence of diffusion in the Anderson tight
binding model for large disorder or low energy},
Commun. Math. Phys. {\bf 88},
  151--184 (1983)

\bibitem{AM}
M. Aizenman and S. Molchanov, {\sl Localization at large disorder and at
extreme energies: an elementary derivation}, Commun.
 Math. Phys. {\bf 157},  245--278  (1993)

\bibitem{RS} I. Rodnianski, W. Schlag, {\sl
Classical and quantum scattering for a class of long range random potentials. }
Int. Math. Res. Not. {\bf 5}  243--300 (2003).

\bibitem{B} J. Bourgain, {\sl Random lattice Schr\"odinger operators
with decaying potential: some higher dimensional phenomena.} Lecture Notes
in Mathematics, Vol. 1807, 70-99 (2003).

\bibitem{Sp1} H. Spohn: {\sl Derivation of the transport equation for
electrons moving through random impurities.}   J. Statist. Phys.{\bf 17}
(1977), no. 6.,
385-412.

\bibitem{EY}  L. Erd\H os and  H.-T. Yau,
{\sl Linear Boltzmann equation as
the weak coupling limit of the random Schr\"odinger equation},
Commun. Pure Appl. Math.
\textbf{LIII}, 667-735, (2000).

\bibitem{Ch} T. Chen, {\sl
Localization Lengths and Boltzmann Limit for the
Anderson Model at Small Disorders in Dimension 3.}
J. Stat. Phys. {\bf 120} (2005), no. 1-2, 279-337.

\end{thebibliography}

\end{document}